\documentclass[%
 reprint,
groupedaddress,
 amsmath,amssymb,
 aps,
]{revtex4-2}

\usepackage[colorlinks=true, citecolor=magenta, linkcolor=blue, urlcolor=blue]{hyperref}
\usepackage{graphicx}
\usepackage{dcolumn}
\usepackage{bm}

\newcommand{\bq}{{\bf q}}

\begin{document}

\preprint{APS/123-QED}

\title{Observation of a new light-induced skyrmion phase in the Mott insulator Cu$_2$OSeO$_3$}

\author{Alexey A. Sapozhnik}
\altaffiliation{The authors have contributed equally}
\author{Benoit Truc}
\altaffiliation{The authors have contributed equally}
\author{Phoebe Tengdin}
\altaffiliation{The authors have contributed equally}
\author{Simone Gargiulo}
\author{Ivan Madan}
\author{Thomas LaGrange}
\author{Fabrizio Carbone}
\altaffiliation{fabrizio.carbone@epfl.ch}
\affiliation{Laboratory for Ultrafast Microscopy and Electron Scattering (LUMES), Institute of Physics, École Polytechnique Fédérale de Lausanne (EPFL), Lausanne, Switzerland.}%

\author{Thomas Sch\"onenberger}
\author{Henrik M. R{\o}nnow}
\affiliation{Laboratory for Quantum Magnetism (LQM), Institute of Physics, École Polytechnique Fédérale de Lausanne (EPFL), Lausanne, Switzerland.}

\author{Arnaud Magrez}
\affiliation{Crystal Growth Facility, Institute of Physics, École Polytechnique Fédérale de Lausanne (EPFL), Lausanne, Switzerland.}



\author{Emil Vi\~nas Bostr\"om}
\affiliation{Max Planck Institute for the Structure and Dynamics of Matter, Hamburg, Germany}%

\author{Angel Rubio}
\affiliation{Max Planck Institute for the Structure and Dynamics of Matter, Hamburg, Germany}%
\affiliation{Center for Computational Quantum Physics (CCQ), The Flatiron Institute, New York, USA.}%

\author{Claudio Verdozzi}
\affiliation{Division of Mathematical Physics and ETSF, Lund University, Lund, Sweden}%

\date{\today}

\begin{abstract}
We report the discovery of a novel skyrmion phase in the multiferroic insulator Cu$_2$OSeO$_3$ for magnetic fields below the equilibrium skyrmion pocket. This phase can be accessed by exciting the sample out of equilibrium with near-infrared (NIR) femtosecond laser pulses but can not be reached by any conventional field cooling protocol. From the strong wavelength dependence of the photocreation process and via spin dynamics simulations, we identify the magnetoelastic effect as the most likely photocreation mechanism. This effect results in a transient modification of the magnetic interaction extending the equilibrium skyrmion pocket to lower magnetic fields. Once created, the skyrmions rearrange and remain stable over a long time, reaching minutes. The presented results are relevant for designing high-efficiency non-volatile data storage based on magnetic skyrmions.
\end{abstract}

\maketitle


\section{\label{sec:level1}Introduction}

Magnetic skyrmions are topologically nontrivial magnetic textures where the spins twist in a vortex-like fashion around the skyrmion core. Their small size and high speed of current-induced motion make them prospective for various spintronics applications \cite{Woo2016, Fert2017, Back2020, Luo2021}. Implementing these concepts requires solving multiple fundamental and technological challenges, such as stabilizing room temperature and zero-field skyrmion phases necessary for practical applications in modern information technology. However, in most of the discovered skyrmion-hosting compounds, skyrmions exist only at low temperatures and require external magnetic fields \cite{Tokura2020}. Searching for topologically nontrivial phases at ambient conditions and exploring the ways for their ultrafast manipulation can lead to a different data storage paradigm, allowing for faster data processing without ohmic losses.

Insulating skyrmion hosting compounds are of high interest due to low Gilbert damping \cite{Stasinopoulos2017}, which allows for studying the propagation of magnons through a skyrmion crystal \cite{Schtte2014, Iwasaki2014} or thermal-gradient induced skyrmion motion \cite{Yu2021}. However, such materials are very rare, including multiferroic Cu$_2$OSeO$_3$ \cite{Seki2012} and Tm$_3$Fe$_5$O$_{12}$ (TmIG), where the topological Hall effect was detected at room temperature in Pt/TmIG heterostructures \cite{Shao2019}. Cu$_2$OSeO$_3$ is the ideal candidate for studying the light-induced effects due to its bulk Dzyaloshinskii–Moriya interaction (DMI) and a rich phase diagram containing various low-temperature magnetic phases \cite{Chacon2018}. The material exhibits a bandgap of 2.5\,eV \cite{Ogawa2015} and a local maximum of absorption around 1.5\,eV corresponding to the transitions between the 3d levels of Cu split by crystal field effects \cite{Versteeg2016}. The equilibrium skyrmion phase in Cu$_2$OSeO$_3$ can be tuned by an external electric field \cite{Okamura2016, Kruchkov2018}, and the electric field-induced creation of skyrmions was recently demonstrated \cite{Huang2018}. The mechanical strain plays an important role in stabilizing skyrmions in Cu$_2$OSeO$_3$, which was evidenced by a significant expansion of the skyrmion phase at high pressures \cite{Levati2016}.

Light stimulation provides a fast and versatile way to control the structural and magnetic properties of the materials \cite{Basov2017}. The coupling between light and the magnetic state of a sample occurs via several mechanisms. These include coupling between the magnetic and the electronic subsystems at an elevated temperature \cite{Koopmans2009}, nonlinear phononics \cite{Afanasiev2021}, or a transient magnetic field generated via the inverse Faraday effect \cite{Kimel2005}. The previous experiments on the photocreation of topological magnetic textures focused on metallic compounds, where the transient heating of the material was identified as the primary microscopic mechanism \cite{Finazzi2013, Berruto2018, Bttner2020}.

In this work, we demonstrate the photoinduced creation of skyrmions by a single NIR femtosecond pulse outside the adiabatically accessible regime. We visualized the skyrmions by the Lorentz transmission electron microscopy (TEM) technique, which provides a high spatial resolution for studying the magnetic materials on the nanoscale \cite{Moeller2020, RubianodaSilva2018, Cao2021}. We successfully generated the skyrmions at low magnetic fields below the equilibrium skyrmion pocket in Cu$_2$OSeO$_3$ by 780-nm and 1200-nm pulses. Considering the low absorption of Cu$_2$OSeO$_3$ at 1200-nm, the reported results are relevant for low-power skyrmion-based memory \cite{Versteeg2016}.

\section{Experimental techniques and methods}
Single crystals of Cu$_2$OSeO$_3$ were grown by chemical vapor transport in a horizontal two-zone furnace. The precursor for the growth was a stoichiometric mixture of CuO and SeO$_2$ sealed in a quartz ampule. The ampule was filled with HCl at a pressure of 100\,mbar, acting as the transport agent. A slab of material with a [111] direction normal to it was cut from a Cu$_2$OSeO$_3$ single crystal and polished to a thickness of 10\,$\mu$m. A TEM lamella was prepared by further thinning down a 5$\times$5\,$\mu$m$^2$ to a thickness of approximately 150\,nm by Ga ions using the focused ion beam (FIB) technique. The thickness of the sample was determined by electron-energy loss spectroscopy (EELS) log-ratio method \cite{Malis1988}.

The measurements were performed in a TEM JEOL 2100HR (acceleration voltage of 200\,kV) equipped with a thermionic gun. The microscope is modified to provide laser light onto the sample \cite{Piazza2013}. The setup was operated at saturation conditions with an electron energy distribution width of ~1\,eV. The images were recorded on a K2 camera (GATAN) in energy-filtered mode, and the width of the energy-selective slit was set to 10\,eV.  Magnetic contrast was achieved using Lorentz transmission electron microscopy (LTEM) with an underfocus of 2\,mm \cite{Phatak2016}.

A Ti:Sapphire regenerative amplifier RAEA HP (KMLabs) was used to produce 35\,fs pulses with a bandwidth centered at 780\,nm at a repetition rate of 4\,kHz. A modified OPA TOPAS (Light Conversion) was utilized for converting them into NIR pulses with a wavelength of 1200\,nm. The duration of the 1200-nm pulses was fixed at 100\,fs, and that of the 780-nm pulses could be changed from 100\,fs to 10\,ps. A set of optical choppers was implemented for reducing the repetition rate down to 4\,Hz necessary for performing single pump pulse experiments enabled by a fast mechanical shutter.

\section{Results}
\begin{figure*}
\includegraphics[width=170mm]{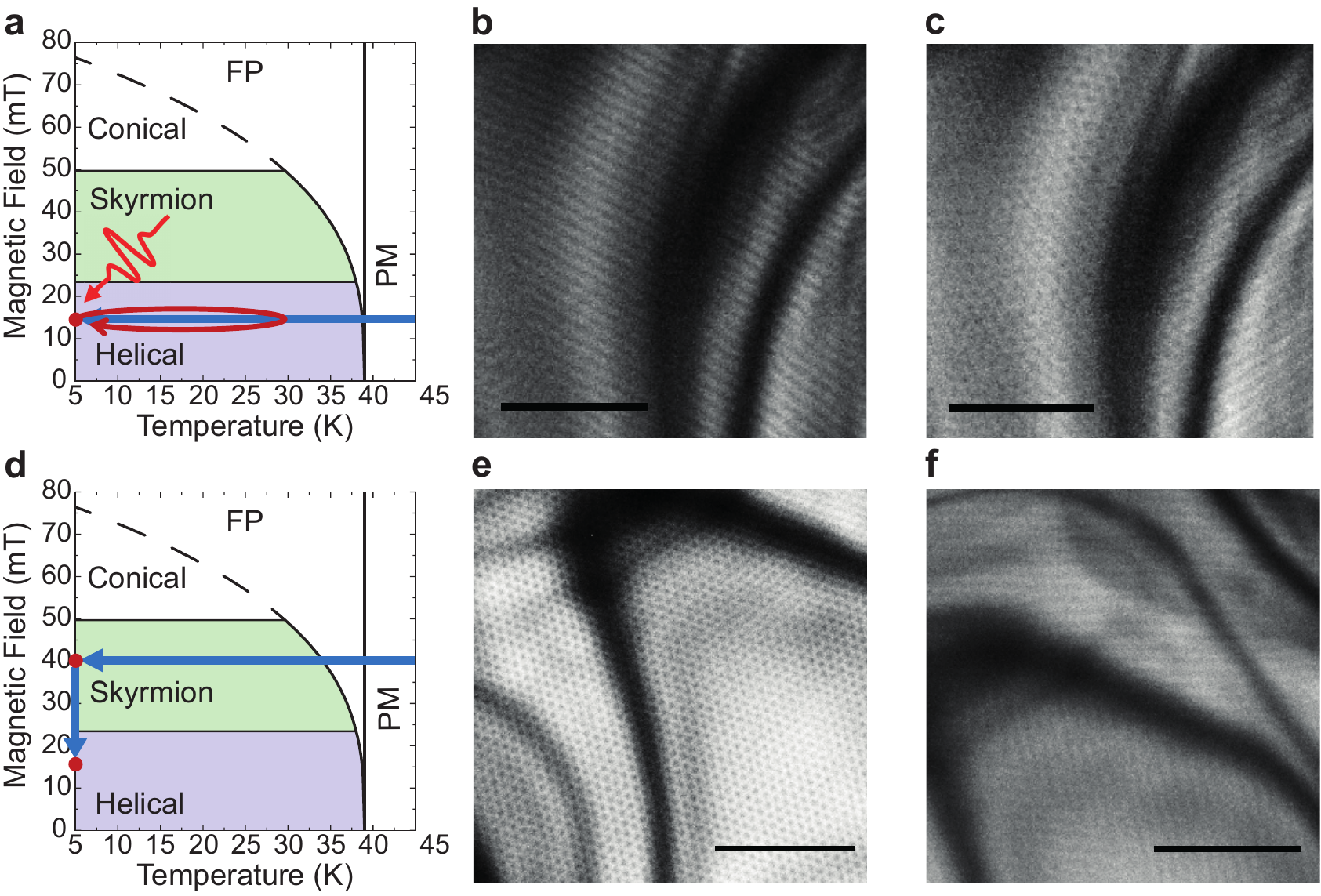}
\caption{\label{fig:PhotocreationAndSlowFC}\textbf{New skyrmion phase below the equilibrium skyrmion pocket in Cu$_2$OSeO$_3$.}\textbf{a} The field cooling phase diagram of the Cu$_2$OSeO$_3$ lamella. FP is the field polarized state, and PM denotes the paramagnetic state. \textbf{b} After FC the sample below the equilibrium skyrmion pocket (14\,mT), only the helical state is visible in the real-space LTEM image measured at 5\,K. \textbf{c} After the arrival of a single 780-nm laser pulse, the magnetic state of the sample contains coexisting skyrmion and helical phases. \textbf{d} Protocol followed to attempt accessing the novel skyrmion phase at low magnetic fields by FC. \textbf{e} The skyrmion phase at 5\,K and 40\,mT generated by field cooling the sample through the Curie temperature. \textbf{f} After decreasing the magnetic field to 14\,mT, the skyrmions disappear and the helical phase emerges. The scale bars in panels \textbf{b}, \textbf{c}, \textbf{e}, \textbf{f} are $1\,\mu$m.}
\end{figure*}

The equilibrium phase diagram of the Cu$_2$OSeO$_3$ lamella was measured by field cooling (FC) the sample from 65\,K at different magnetic fields and is presented in Fig.~\ref{fig:PhotocreationAndSlowFC}a. The cooling rate was $\sim$1\,K/sec. Field cooling in magnetic fields lower than 24\,mT results in the appearance of a helical phase below the ordering temperature $T_C$. Magnetic fields exceeding 50\,mT correspond to the conical or field polarized (FP) states that are indistinguishable in the LTEM images. The boundary between these two phases is indicated by a dashed line in Fig.~\ref{fig:PhotocreationAndSlowFC}a and d. The measured $T_C$ of 40 K is lower than the value of 59\,K reported earlier \cite{Seki2012}. This discrepancy might be explained by the difference in the actual sample temperature and the temperature measured by the thermocouple of the sample holder or by the fact that we studied a thin lamella (150\,nm) that may have a different $T_C$ and phase diagram than the bulk crystal measured in Ref. \cite{Seki2012}.

The skyrmion photocreation experiment below the equilibrium skyrmion pocket was conducted according to the protocol shown in Fig.~\ref{fig:PhotocreationAndSlowFC}a. First, the sample was cooled from 65\,K (above the $T_C$) in a field of 14\,mT, which is the remanent field of the TEM objective lens, following the blue arrow. The initial state of the sample after FC is helical (Fig.~\ref{fig:PhotocreationAndSlowFC}b). After reaching a temperature of 5\,K, the sample was irradiated with a single 780-nm femtosecond pulse resulting in the formation of a skyrmion lattice (Fig.~\ref{fig:PhotocreationAndSlowFC}c). The absorbed fluence in this experiment was 15\,mJ/cm$^2$.

We tested the possibility of creating the low-field skyrmion phase shown in Fig.~\ref{fig:PhotocreationAndSlowFC}c by slow field cooling. Fig.~\ref{fig:PhotocreationAndSlowFC}d indicates the path followed within the phase diagram. First, the sample is cooled in a field of 40\,mT from 65\,K to 5\,K at a cooling rate of $\sim$1\,K/sec, resulting in the formation of a skyrmion lattice (Fig.~\ref{fig:PhotocreationAndSlowFC}e). However, after decreasing the magnetic field to 14\,mT, the skyrmion lattice transforms into helices (Fig.~\ref{fig:PhotocreationAndSlowFC}f). Thus, the photoinduced skyrmion phase in the Cu$_2$OSeO$_3$ lamella at 14\,mT and 5\,K is a manifestation of a unique magnetic phase, which can be accessed only via photoexcitation of the sample.

\begin{figure*}
\includegraphics[width=170mm]{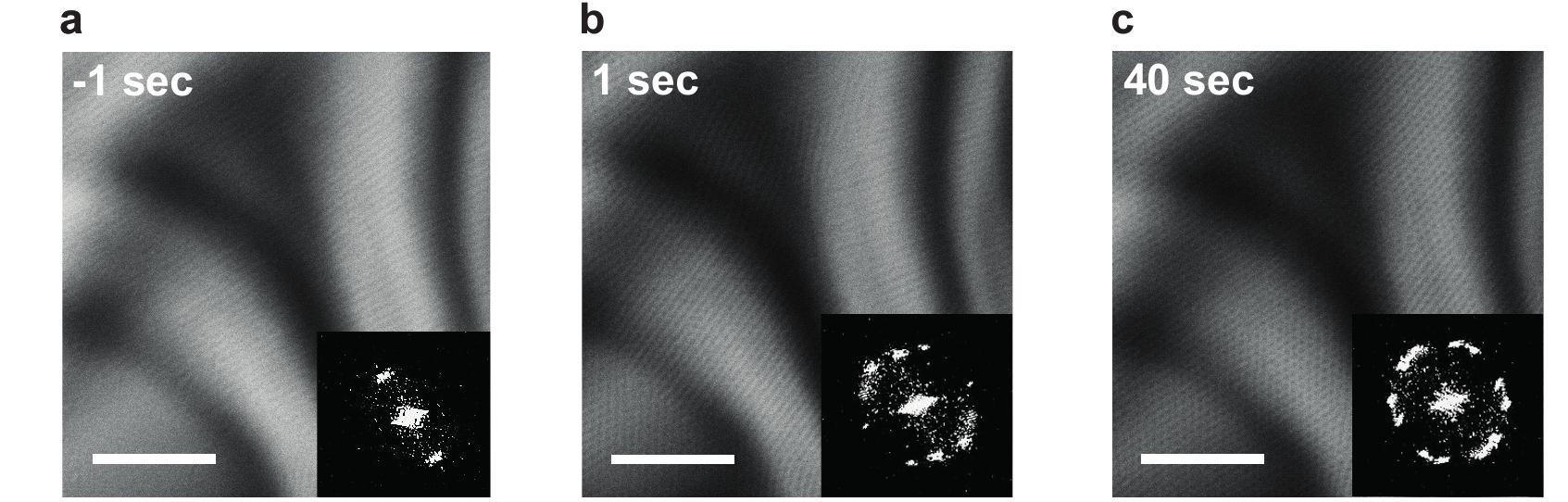}
\caption{\label{fig:Stability}\textbf{The stability of the photocreated skyrmion phase.} The LTEM images of the Cu$_2$OSeO$_3$ lamella at 5\,K and 14\,mT \textbf{a} one second before, \textbf{b} one second after, and \textbf{c} 40 seconds after the irradiation of the sample by an optical pulse. The insets show the Fourier Transform (FT) patterns calculated over the corresponding images. The scale bar is 1\,um.}
\end{figure*}

The discovered low-field skyrmion phase exhibits a long lifetime exceeding several tens of seconds (Fig.~\ref{fig:Stability}). We have also observed the skyrmions present in the sample minutes after the excitation (not shown). The helical and the skyrmion domains coexist in the sample for a few seconds after the optical excitation (Fig.~\ref{fig:Stability}b), and on a longer time scale, the skyrmion lattice expands prevailing over the helical domains (Fig.~\ref{fig:Stability}c).

\begin{figure*}
\centerline{\includegraphics[width=170mm]{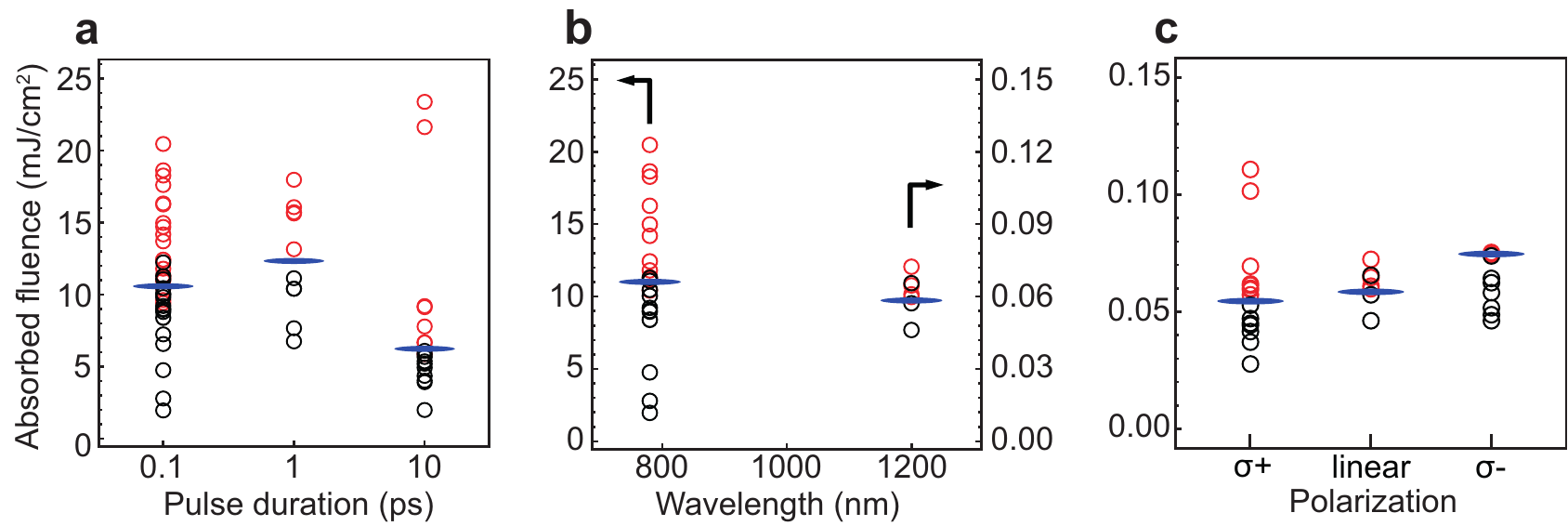}}
\caption{\label{fig:PulseDurationAndWavelength}\textbf{Absorbed fluence threshold for the photoinduced generation of skyrmions with NIR photons.} \textbf{a} The pulse duration dependence of the skyrmion photocreation in Cu$_2$OSeO$_3$ at 5\,K and 14\,mT for a 780-nm single-pulse photoexcitation. \textbf{b} Comparison of the photocreation threshold between 780-nm and 1200-nm optical excitation. \textbf{c} The photocreation of skyrmions with single femtosecond pulses having a wavelength of 1200\,nm for different polarizations. The vertical axis indicates the absorbed fluence. The red and black circles display the final state of the sample exhibiting skyrmions and helices, respectively. The blue horizontal lines indicate the skyrmion generation threshold. The vertical axes in all panels correspond to the absorbed fluence in the thin part of the sample.}
\end{figure*}

We studied the pulse duration dependence of the skyrmion photocreation process at 5\,K and a field of 14\,mT. The final magnetic states of the Cu$_2$OSeO$_3$ lamella are reported in Fig.~\ref{fig:PulseDurationAndWavelength}a for the excitation with a single 780-nm pulse. Before each measurement, the sample was reset to the single-domain helical state by briefly applying an out-of-plane magnetic field of 1\,T. Note that resetting by going above the Curie temperature and following field cooling did not result in any change. The vertical axis corresponds to the fluence absorbed in the thin 150-nm part of the sample (see Supplementary Information). The sample remains in the helical state for lower fluences, as indicated by the black circles. Occasionally, a reorientation of the helix wave vector Q was observed after the arrival of a pulse, but no skyrmions were created. A skyrmion lattice was generated as higher fluences, corresponding to the red circles. The skyrmion generation threshold marked by the blue lines shows only weak dependence on the pulse duration.

Tuning the photon energy below the crystal field excitation regime allows for achieving an outstanding efficiency in generating the skyrmions as demonstrated in Fig.~\ref{fig:PulseDurationAndWavelength}b. The absorbed threshold fluence required for generating the skyrmions with 780-nm pulses at 5\,K and 14\,mT is 11\,mJ/cm$^2$, and for 1200-nm photons, this value reduces to 0.06\,mJ/cm$^2$. Although the absolute values varies slightly among the samples tested, we reproduced this result in four different samples. To the best of our knowledge, it is a record-low fluence necessary for generating skyrmions in a magnetic material. The fluence threshold shows only a weak polarization dependence for a wavelength of 1200\,nm (Fig.~\ref{fig:PulseDurationAndWavelength}c). This behavior is expected from the cubic symmetry of the material point group and is consistent with the simulations discussed below.

\section{Discussion}
The low-temperature skyrmion phase identified in our experiments can only be accessed by laser excitation (Fig.~\ref{fig:PhotocreationAndSlowFC}c) and is absent under adiabatic field cooling of the sample (Fig.~\ref{fig:PhotocreationAndSlowFC}f). It indicates that the photocreation of skyrmions in this regime cannot be explained solely by the transient heating of the sample and that other non-thermal effects play a central role. To gain further insight into the microscopic processes underlying skyrmion photocreation in Cu$_2$OSeO$_3$, we performed extensive spin dynamics simulations. Due to the multiferroic nature of the material, there is a large number of mechanisms by which the laser electric field can affect magnetization. Since our measurements were performed at wavelengths inside the bulk band gap and low to moderate fluences, real (as opposed to virtual) electronic excitations can be assumed negligible. In addition, as indicated by Fig.~\ref{fig:PulseDurationAndWavelength}b, a much lower fluence is needed to create skyrmions for $\lambda$ = 1200\,nm than for $\lambda$ = 780\,nm. It is likely due to the electronic crystal field excitations present at shorter wavelengths \cite{Versteeg2016}, which hinder energy from reaching the magnetic subsystem and are detrimental to skyrmion creation. 

\begin{figure*}
 \includegraphics[width=\textwidth]{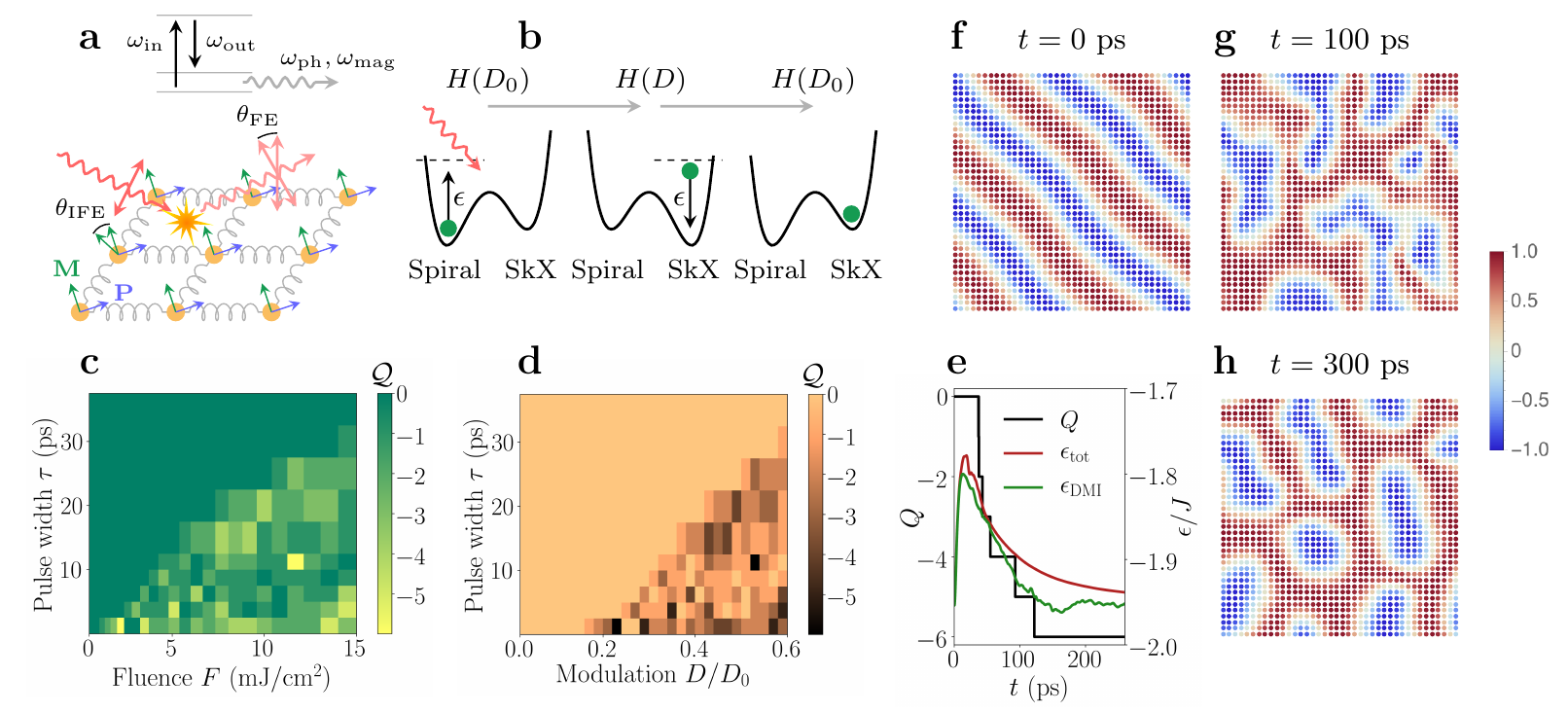}
 \caption{{\bf Microscopic mechanism of skyrmion photocreation.} {\bf a,} Schematic of the light-matter interaction mechanisms contributing to skyrmion photocreation in Cu$_2$OSeO$_3$. Top panel: Raman scattering processes involving virtual electronic states lead to emission of magnons and phonons. Main panel: The inverse Faraday effect and a direct magneto-electric coupling to the polarization (blue arrows) leads to a change in laser polarization and a torque on the magnetic moments (green arrows). {\bf b,} Heuristic rendering of the mechanism of skyrmion photocreation: $(1)$ Raman and magneto-electric processes lead to excitation of the coupled spin-phonon system and a change in the free energy landscape. $(2)$ The system relaxes into a quasi-stationary state defined by the photon-modulated Hamiltonian $(3)$ The system returns to equilibrium trapped in the meta-stable skyrmion crystal state. {\bf c,} Non-equilibrium magnetic phase diagram as a function of laser fluence and pulse duration. The areas with a non-zero value of the topological charge $\mathcal{Q}$ correspond to skyrmion photocreation. {\bf d,} Topological charge $Q$ as a function of phonon-modulated Dzyaloshinskii-Moriya interaction $D$. {\bf e,} Time-dependence of the topological charge and the total energy  per spin during skyrmion creation. {\bf f-h,} Instantaneous spin configurations at different times $t$ for the pulse duration $\tau = 10\,ps$ and a fluence $F = 10$\,mJ/cm$^{2}$. The equilibrium spin parameters of the normalized spin model are taken as $J = 48.2$\,meV, $D = 10.8$ and $B = 0.63$\,meV (corresponding to $B = 20.7$\,mT), in accordance with literature. The strength of the light-matter coupling constants are taken as $g_{\rm R} = 0.1J$, $g_{\rm IFE} = 0$, $g_{\rm m-el} = 0.01J$ and $g_{\rm m-ph} = 0.5D$ for a laser electric field of $E = 10^9$\,V/m.}
 \label{fig:theory}
\end{figure*}

As illustrated in Fig.~\ref{fig:theory}a, the possible light-matter coupling mechanisms are the Raman excitation of phonons and magnons, an effective magnetic field generated by the inverse Faraday effect, and direct magnetoelectric coupling to the spontaneous polarization~\cite{Yang2012,Seki2012b,Liu2013}. The magnetoelectric effect, originating from the coupling between the laser electric field and the electronic polarization of Cu$_2$OSeO$_3$, is proportional to the amplitude of the electric field~\cite{Huang2018}. Thus, the fluence threshold is expected to show a strong dependence on the pulse duration and a crystal orientation dependence that we did not observe. In contrast, our experimental data only exhibits a weak dependence of the fluence threshold on the pulse duration (Fig.~\ref{fig:PulseDurationAndWavelength}a). Thus, the magnetoelectric coupling is likely not the main mechanism driving the observed photocreation below the equilibrium skyrmion pocket. The inverse Faraday effect is also expected to play a small role in the photocreation process due to the weak polarization dependence for 1200-nm light~\cite{Kimel2005}.

In contrast, static mechanical strain is known to modify the shape of the magnetic phase diagram of Cu$_2$OSeO$_3$ and similar materials. An increase of the $T_C$ and an expansion of the skyrmion pocket were demonstrated in a bulk crystal of Cu$_2$OSeO$_3$ under compressive stress~\cite{Levati2016}. Moreover, a negative uniaxial strain can shift the equilibrium skyrmion pocket to lower magnetic fields via magnetoelastic coupling~\cite{Wang2019}, and mechanical strain can modify the Dzyaloshinskii-Moriya interaction (DMI) constant of a skyrmion hosting compound~\cite{Ba2021}, or both the DMI constant and the anisotropy constant~\cite{Feng2021}. Since a modest strain of $0.3$ \% can induce a modulation of the DMI of up to 20 \%~\cite{Shibata2015,Deng2020}, a transient strain mediated by long-wavelength acoustic phonons is expected to have a significant impact on the DMI~\cite{Nomura2019}.

To make the above arguments quantitative we consider a time-dependent interaction Hamiltonian $H_I(t)$ accounting for all of the discussed mechanisms. This Hamiltonian is
\begin{align}\label{eq:hamiltonian}
 H_I(t) &= \sum_{\langle ij\rangle} \left[ J_{ij}(t) {\bf m}_i \cdot {\bf m}_j + {\bf D}_{ij}(t) \cdot ({\bf m}_i \times {\bf m}_j) \right] \\
   &+ \sum_i \left[{\bf B}(t) \cdot {\bf m}_i - {\bf E}(t) \cdot {\bf P}_i \right], \nonumber
\end{align}
where $J_{ij}$ is the exchange interaction between magnetic moments ${\bf m}_i$ and ${\bf m}_j$, ${\bf D}_{ij}$ is the DMI, ${\bf B}$ and ${\bf E}$ are the external magnetic and electric fields, and ${\bf P}_i$ the electronic polarization. Each microscopic process is associated with a characteristic energy scale, denoted by $g_{\rm R}$ for the magnon Raman process, $g_{\rm IFE}$ for the inverse Faraday effect, $g_{\rm m-el}$ for the magneto-electric coupling, and $g_{\rm m-ph}$ for the magneto-phonon coupling. Further, each mechanism only affects a given term in Eq.~(\ref{eq:hamiltonian}), such that the exchange depends on $g_{\rm R}$, the DMI on $g_{\rm m-ph}$, the magnetic field on $g_{\rm IFE}$ and the polarization on $g_{\rm m-el}$. A detailed discussion of the values of these energies is provided in the Supplemental Material.

To account for the time-dependence of the magnetic interactions, we note that the magnon Raman process, the inverse Faraday effect, and the magneto-electric effect only modify the spin parameters during the action of the pulse (assumed to be a Gaussian of width $\tau$). In contrast, the phonon modulation of the DMI is expected to persist for as long as there are phonons present in the system. The time-dependent part of the DMI is therefore assumed to have an onset time given by the pulse width $\tau$, and an exponential decay set by the phonon lifetime $\tau_{\rm ph}$. This time dependence is described by a log-normal function as further discussed in the Supplemental Material.

To describe the skyrmion photocreation process we simulated the time-evolution governed by Eq.~(\ref{eq:hamiltonian}) following laser excitation (see Supplemental Material for a discussion of the equilibrium spin parameters and phase diagram as well as the spin equations of motion). In line with our experiments, the magnetic field was chosen such that the system is initially in the helical state and close to the phase boundary to the conical state. By exploring a significant portion of the parameter space defined by Eq.~(\ref{eq:hamiltonian}), the dominant mechanism leading to skyrmion photocreation in Cu$_2$OSeO$_3$ was identified as the transient modulation of the DMI by long-wavelength acoustic phonons. This identification is in line with several previous studies that have found a strong dependence of the Dzyaloshinskii-Moriya interaction (DMI) on strain~\cite{Shibata2015,Deng2020} as well as on a dynamical coupling to acoustic phonons~\cite{Nomura2019}.

In agreement with our experimental results, our simulations predict a skyrmions phase to appear upon laser irradiation for magnetic parameters below the equilibrium skyrmion pocket. Heuristically, the skyrmion photocreation process as emerging from our simulations can be understood as follows (see Fig~\ref{fig:theory}b). The coupled spin-phonon system is excited into a non-equilibrium state through Raman and magneto-electric processes. Simultaneously, the modulation of the DMI by acoustic phonons leads to a change in the free energy landscape, allowing the system to relax into a quasi-stationary state defined by the instantaneous phonon-modulated Hamiltonian. At a time-scale set by the phonon lifetime $\tau_{\rm ph}$ the free energy landscape returns to its original form, while the magnetic system stays trapped in the meta-stable skyrmion crystal state.

To substantiate this picture we show in Fig.~\ref{fig:theory}c the non-equilibrium magnetic phase diagram as a function of pulse duration and laser fluence, and shows the topological charge $\mathcal{Q}$ of magnetic state at the final time of our simulations (corresponding to about $300$\,ps). The topological charge $\mathcal{Q} = 0$ in the helical or ferromagnetic state, and becomes $\mathcal{Q} = -1$ and $1$ for skyrmions and antiskyrmions, respectively. Thus the topological charge counts skyrmions and antiskyrmions and is non-zero only when a net imbalance of such excitations is present. Our simulation shows that photoexcitation strongly favors skyrmion creation, and thus $\mathcal{Q}$ in Fig.~\ref{fig:theory}c is equal to the total skyrmion number $N_{\rm sk}$. Clearly, above a threshold fluence of $F \approx 1$\,mJ/cm$^2$, the non-equilibrium steady state changes character from a helical to a skyrmion crystal state. Similarly Fig.~\ref{fig:theory}d shows the topological charge $\mathcal{Q}$ as a function of DMI modulation.

We note that Fig.~\ref{fig:theory}c predicts an approximately linear relation between the threshold fluence and the pulse duration. For this to be consistent with Fig.~\ref{fig:PulseDurationAndWavelength}a, we have to assume that the spins are insensitive to processes on time-scales shorter than about $1$ ps. More precisely, assuming that the laser energy is transferred to the acoustic phonons faster than some excitation time $\tau_{\rm exc}$, and that this time is shorter than the characteristic magnetic time-scale $\tau_{\rm spin}$, the resulting spin dynamics is expected to be independent of pulse duration for $\tau < \tau_{\rm exc}$.

To assess the stability of the non-equilibrium skyrmion state, we further investigated the time-evolution of the magnetic system. Fig.~\ref{fig:theory}f-h shows the instantaneous magnetization ${\bf m}_i(t)$ at a number of different times $t$, for a pulse length $\tau$ leading to a meta-stable skyrmion state. As seen from Fig.~\ref{fig:theory}e, the topological charge changes during the initial part of the excitation and relaxation process, which can be identified from total energy per spin $\epsilon_{\rm tot} = E/N$. However, after a time $t = 150$\,ps, corresponding roughly to the phonon lifetime $\tau_{\rm ph}$, the topological charge is constant. Within the spin model the skyrmion state remains stable indefinitely, since additional energy would have to be supplied to bring the system back into the helical ground state. This is in good agreement with the experimental finding of a meta-stable skyrmion state surviving for several minutes.

\section{Conclusion}
In conclusion, we demonstrated the possibility of generating skyrmions in Cu$_2$OSeO$_3$ at low magnetic fields below the equilibrium skyrmion pocket by NIR femtosecond laser pulses. Supported by the wavelength dependence and spin dynamics calculations, we claim that the irradiation of the sample results in the triggering of low-energy phonons which transiently change the DMI. As a consequence, these effects modify the free energy landscape of the material and enable the transformation of the magnetic state into the skyrmion lattice. The lifetime of the low-field skyrmion phase after the light-induced generation exceeds minutes, which is important for future magnetic data storage. The threshold fluences for the skyrmion photocreation at 5\,K and 14\,mT are 11\,mJ/cm$^2$ and 0.06\,mJ/cm$^2$ for 780\,nm and 1200\,nm, respectively. The latter is the lowest reported fluence required for generating skyrmions. Thus, our experiment marks a milestone in the development of energy-efficient skyrmion-based spintronics devices.
\\

\begin{acknowledgments}
We acknowledge support from the ERC consolidator grant ISCQuM and SNSF via sinergia nanoskyrmionics grant 171003. The authors would like to gratefully acknowledge Prof. A. Rosch, Dr. N. del Ser, and Prof. J. Zang for helpful discussions and the comments related the theory part of the manuscript.
\end{acknowledgments}

\bibliography{apssamp}

\clearpage

\section*{Supplemental Methods}

\subsection*{Calculation of the absorbed fluence}
To determine the absorbed fluence, we consider a sample thickness $t = 150$ nm and an absorption length of either $l_{780\,\rm nm} = 1.5$ $\mu$m and $l_{1200\,\rm nm} = 30$ $\mu$m. Since $t \ll l_{780\,\rm nm}$, $l_{1200\,\rm nm}$, the absorbed fluence fraction can be approximated as $t/l_{\lambda}$. Thus, 10\,\% and 0.5\,\% of the incident fluence are absorbed at 780\,nm and 1200\,nm, respectively.

\subsection*{Equilibrium spin Hamiltonian}
The magnetic structure of Cu$_2$OSeO$_3$ consists at the microscopic level of 16 Cu ions per unit cell, each carrying a magnetic moment of $|{\bf s}_i| \approx \hbar/2$. However, due a hierarchy of magnetic interaction strengths~\cite{Janson2014}, the four spins on each pyramid of the pyrochlore lattice bind together to form effective magnetic moments of size $|{\bf S}_i| = \hbar$ living on a trillium lattice. After an additional coarse graining step, valid for magnetic structures where the magnetization is constant over distances comparable with the lattice size ($\nabla\cdot{\bf M} \ll a$), the effective spin Hamiltonian in equilibrium is
\begin{align}\label{H_eq}
 H_0 &= \sum_{\langle ij\rangle} \left[ J\, {\bf m}_i \cdot {\bf m}_j + {\bf D}_{ij} \cdot ({\bf m}_i \times {\bf m}_j) \right] \\
   &+ \sum_i {\bf B} \cdot {\bf m}_i. \nonumber
\end{align}
For Cu$_2$OSeO$_3$ the skyrmion radius is $r = 50.89$ nm, compared to the lattice parameter $a = 8.91$ {\AA}, such that the coarse graining procedure is justified. To describe the magnetization dynamics we employ a three-dimensional square lattice with $40\times 40\times 10$ lattice points. The magnetic moments are normalized to $|{\bf m}_i| = 1$, and all interaction parameters are measured in units of the exchange interaction $J_0 = 48.2$ meV. This gives the effective parameters $J = 1$, $D = 0.224$ and $B = 0.013$, corresponding to $20.7$ mT, in agreement with previous work~\cite{Janson2014}.


\subsection*{Simulated annealing and Metropolis Monte Carlo}
The equilibrium magnetic phase diagram is found by simulated annealing down to a target temperature $k_BT/J_0 = 0.02$ using the Metropolis Monte Carlo algorithm~\cite{Bostrom19}, corresponding to $T \approx 1$ K. To minimize stochastic effects in the phase diagram and the subsequent dynamics, each step involves 2000 thermalization sweeps followed by an average over 2000 Monte Carlo realizations with 40 sweeps each. In agreement with previous work~\cite{Buhrandt2013}, we find four competing equilibrium phases: a ferromagnetic phase, a helical spiral phase, a conical phase and a skyrmion crystal (SkX) phase.


\subsection*{Equations of motion}
The magnetization dynamics is governed by the Landau-Lifshitz-Gilbert (LLG) equation~\cite{Lakshmanan11,Baryakhtar15}, which in the present case reads
\begin{align}
 \frac{\partial {\bf m}_i}{\partial t} &= - \gamma {\bf m}_i \times \frac{\delta H}{\delta {\bf m}_i} - \lambda {\bf m}_i \times \left( {\bf m}_i \times \frac{\delta H}{\delta {\bf m}_i} \right).
\end{align}
Here the effective magnetic field acting on magnetic moment ${\bf m}_i$ is given by the functional derivative of the total Hamiltonian $H$ with respect to ${\bf m}_i$. The parameters $\gamma = 1/(1+\alpha^2)$ and $\lambda = \alpha/(1+\alpha^2)$ take into account the phenomenological damping constant $\alpha$, which for Cu$_2$OSeO$_3$ is on the order of $10^{-4}$. The LLG equation is solved by geometric Depondt-Mertens algorithm~\cite{Depondt2009}.


\subsection*{Light-matter coupling}
The dominant light-matter coupling mechanisms considered here are Raman excitation of phonons and magnons, an effective magnetic field generated by the inverse Faraday effect, and a direct magneto-electric coupling via the spontaneous polarization~\cite{Yang2012,Seki2012,Liu2013}. Several studies have found a strong dependence of the Dzyaloshinskii-Moriya interaction (DMI) on strain~\cite{Shibata2015,Deng2020} as well as on a dynamical coupling to acoustic phonons~\cite{Nomura2019}. To describe such mechanisms we consider the total Hamiltonian $H(t) = H_0 +H_I(t)$, with the time-dependent interaction Hamiltonian
\begin{align}
 H_I(t) &= \sum_{\langle ij\rangle} \left[ J_{ij}(t) {\bf m}_i \cdot {\bf m}_j + {\bf D}_{ij}(t) \cdot ({\bf m}_i \times {\bf m}_j) \right] \\
   &+ \sum_i \left[{\bf B}(t) \cdot {\bf m}_i - {\bf E}(t) \cdot {\bf P}_i \right], \nonumber
\end{align}
The exchange interaction $J_{ij} = g_{\rm R}(t) ({\bf e}_{\rm sc} \cdot {\bf d}_{ij}) ({\bf e}_{\rm in} \cdot {\bf d}_{ij})$ is modified to take into account magnon Raman processes, the inverse Faraday effect generates an effective field ${\bf B}(t) = g_{\rm IFE} {\bf E}^{*}(t) \times {\bf E}(t)$, and the magneto-electric effect is described by a coupling to the polarization ${\bf P}_i = g_{\rm m-el} (S_i^y S_i^z, S_i^z S_i^x, S_i^x S_i^y)$. Assuming a uniform excitation of acoustic phonons with momenta ${\bf k} \approx 0$ gives an isotropic modification of the DMI strength ${\bf D}(t) = {\bf D} (1 - g_{\rm m-ph}(t))$, where $g_{\rm m-ph}(t)$ is proportional to the time-dependent average phonon amplitude. Since the phonon dynamics of Cu$_3$OSeO$_2$ is very complex, we here use a phenomenological description of $g_{\rm m-ph}(t)$ as a log-normal function, with an onset determined by the laser electric field and a decay related to the phonon lifetime $\tau_{\rm ph}$. The laser electric field is described by a normalized Gaussian envelope of width $\sigma$ and peak time $\tau$.


\subsection*{Time-dependence of the magnetic parameters}
The excitation mechanisms discussed above are associated with different characteristic time-scales. In particular, both the magnon Raman processes, inverse Faraday effect and magneto-electric effects are impulsive in the sense that they only are present during the action of the laser pulse. In contrast, the magnon-phonon coupling is expected to persist for as long as there are phonons in the system. Since the magnetic moments have a characteristic time-scale of about $1$ ps, while a single optical cycle of the laser is around $1$ fs, the magnetic moments are assumed to only respond to the average field given by the pulse envelope. 

Under these assumptions, the first three mechanisms satisfy $g_i(t) = g_i f(t)$ for $g_i \in \{g_{\rm R}, g_{\rm IFE}, g_{\rm m-el}\}$ and can be described by a Gaussian time-dependence of the form
\begin{align}
 f(t) = \frac{1}{\tau\sqrt{2\pi}} \exp[-(t-t_0)^2/(2\tau^2)],
\end{align}
where $\tau$ is the pulse width and $t_0$ the time of peak intensity. In contrast, the magnon-phonon coupling is modeled by the log-normal function
\begin{align}
 g_{\rm m-ph}(t) = \frac{g_{\rm m-ph}}{t\sigma\sqrt{2\pi}} \exp[-(\ln t-\mu)^2/(2\sigma^2)]
\end{align}
Here the peak time is given by $t_0 = e^{\mu-\sigma^2}$, and the effective width (or skewness) by $\tau = (e^{\sigma^2} + 2) \sqrt{e^{\sigma^2} - 1}$. The width is assumed to be related to the phonon lifetime, which is assumed to be of the order of $100$ ps. A few examples of the log-normal function for $\sigma = 1$ and different $t_0$ is given in Fig.~\ref{fig:lognormal}.

\setcounter{figure}{0}
\renewcommand{\figurename}{FIG.}
\renewcommand{\thefigure}{S\arabic{figure}}

\begin{figure}
 \includegraphics[width=0.8\columnwidth]{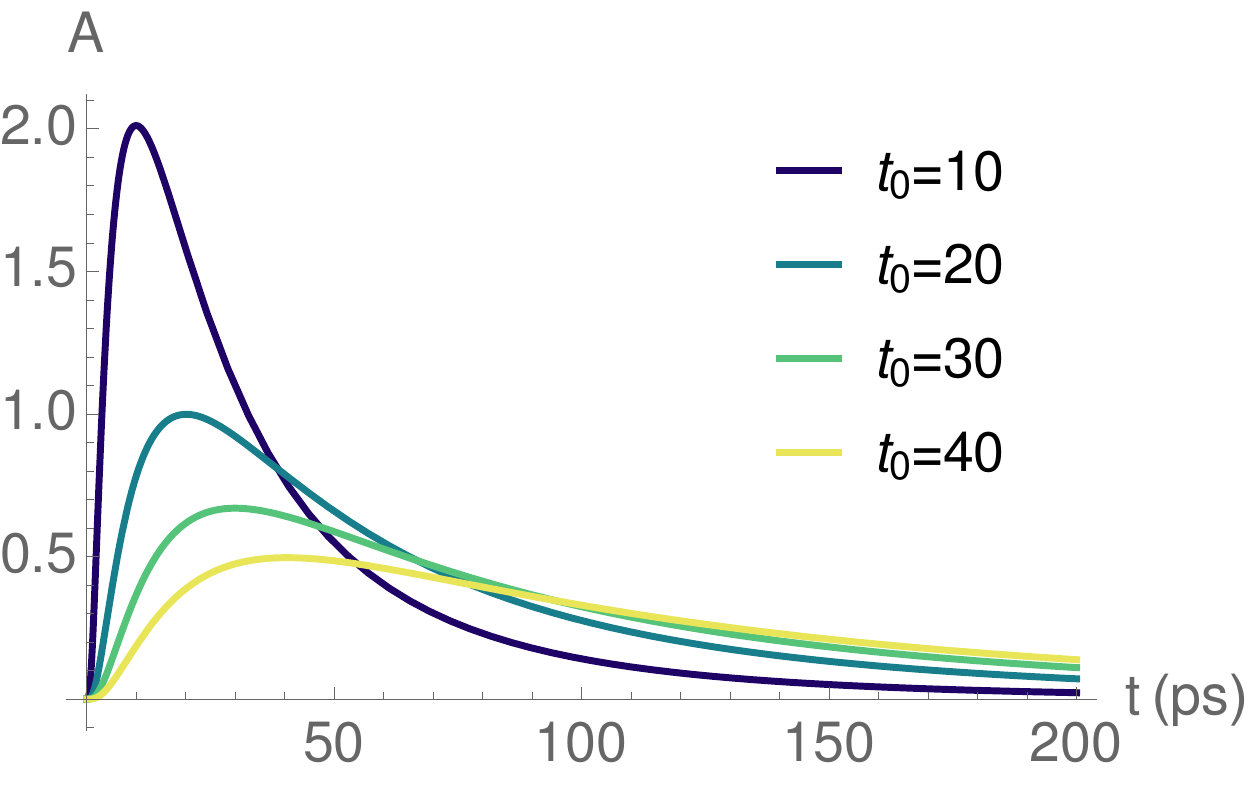}
 \caption{{\bf Time-dependence of magnon-phonon interaction.} Examples of log-normal functions with peak time $t_0$ as used to model time-dependence of the magnon-phonon interaction.}
 \label{fig:lognormal}
\end{figure}


\subsection{Definition of the topological charge}
To quantify the topology of the spin texture we use the lattice topological charge $Q$ defined by~\cite{Berg81} 
\begin{align}
Q = \frac{1}{4\pi}\sum_\Delta \Omega_\Delta.
\end{align}
In this definition the lattice is triangulated and $\Omega_\Delta$ corresponds to the signed area of the spherical triangle spanned by three neighboring spins, given by~\cite{Berg81}
\begin{align}
 \exp&(i\Omega_\Delta/2) = \frac{1}{\rho}(1 + {\bf m}_i \cdot {\bf m}_j+ {\bf m}_j \cdot {\bf m}_k + {\bf m}_k \cdot {\bf m}_i \\
 &\hspace*{2.3cm}+ i\eta_{ijk} {\bf m}_i \cdot [{\bf m}_j \times {\bf m}_k]) \nonumber \\
 \rho &= \sqrt{2(1+{\bf m}_i \cdot {\bf m}_j)(1+{\bf m}_j \cdot {\bf m}_k)(1+{\bf m}_k \cdot {\bf m}_i)}, \nonumber
\end{align}
where $\eta_{ijk} = +1$ $(-1)$ if the path $i\to j\to k\to i$ is positively (negatively) oriented. The surface area $\Omega_\Delta$ is well-defined everywhere except at the zero-measure set ${\bf m}_i\cdot ({\bf m}_j \times {\bf m}_k) = 0$ and $1 + {\bf m}_i \cdot {\bf m}_j+ {\bf m}_j \cdot {\bf m}_k + {\bf m}_k \cdot {\bf m}_i < 0$, where $\exp(i\Omega_\Delta/2)$ has a branch cut. 

The topological charge is a compact, convenient indicator of the presence of a non-trivial spin texture: For a single skyrmion $Q = -1$, for a single antiskyrmion $Q = 1$, and for a cluster of skyrmions and antiskyrmions $Q = \sum_i Q_i$ with $Q_i$ their individual charges.


\section*{Supplemental Note: Estimated magnitude of light-matter couplings}
As discussed in detail in the following sections, the magnon Raman coupling gives a modulation $\sim 10$ \% of the equilibrium exchange, the inverse Faraday effect is likely negligible, the magneto-electric effect gives a contribution of around $\sim 1$ \% of the equilibrium exchange, while the spin-phonon coupling can give a modulation on the order of $\sim 50$ \% of the equilibrium DMI.

\subsection*{Magnon Raman processes}
An isotropic light-matter interaction arises due to the coupling of the laser to the charge of the electrons underlying the magnetic moments. For spin-$1/2$ systems, this coupling can be derived by considering a half-filled Mott insulator subject to an external electric field. The weak-field limit of this coupling reproduces the Raman vertex derived by Fleury and Loudon from general symmetry arguments~\cite{Shastry90,Fleury67,Fleury68}, and is described by the Raman Hamiltonian~\cite{VinasBostrom2021}
\begin{align}\label{eq:raman_ham}
 H_R &= \sum_{qq'} R_{\bf qq'} a_{q'}^\dagger a_{q} \sum_{\langle ij\rangle} g_{ijqq'} {\bf m}_i \cdot {\bf m}_j \\
 &= \sum_{\langle ij\rangle} J_{ij} {\bf m}_i \cdot {\bf m}_j. \nonumber
\end{align}
Here $R_{\bf qq'} = J (ea/\hbar)^2 \gamma_\bq \gamma_{\bf q'}$ is the strength of the Raman coupling, $e$ is the electron charge, $a$ the lattice parameter, $\hbar$ Planck's constant, and $J$ is the equilibrium exchange interaction. The function $\gamma_\bq$ describes the strength of the one-photon vector potential, and the geometric factor $g_{ijqq'} = (\hat{\bf e}_q^{*} \cdot {\bf d}_{ij})(\hat{\bf e}_{q'} \cdot {\bf d}_{ij})$ encodes the underlying virtual electronic processes. To simplify the notation we have defined $q \equiv \{\bq,s\}$ with $s$ the polarization.

The leading order term of the Raman Hamiltonian describes a two-photon two-magnon process. To assess the strength of the light-matter coupling, we note that $\lambda_{R,\bq} = R_{\bf qq'} n_\bq$ where $n_\bq$ is the number of photons in the incident field. Using the fact that $n_\bq = IV/(\hbar\omega_\bq c)$ with the intensity $I = (cn\epsilon_0/2) E^2$, we have $\lambda_{R,\bq} = (ea/\hbar\omega_\bq)^2 (n/4) E^2$. Assuming that $a = 5$ {\AA}, $E = 10^9$ Vm$^{-1}$ and $n = 2.40$, we find $\lambda_R = 0.16$ with and characteristic energy scale $g_{\rm R} = J\lambda_R = 7.7$ meV.


\subsection*{Inverse Faraday effect}
For a system with non-zero magnetization the dielectric tensor acquires nonzero off-diagonal elements and can be written as $\epsilon_{ij}({\bf M}) = \epsilon_0(\epsilon_r \delta_{ij} - if \epsilon_{ijk} M_k)$. Here $\epsilon_0$ is the vacuum permittivity, $\epsilon_r$ the relative permittivity, $f$ is a small parameter related to the Faraday angle $\theta_F$ discussed below, and $\delta_{ij}$ and $\epsilon_{ijk}$ are the Kronecker and Levi-Cevita tensors, respectively. Calculating the interaction energy in a volume $V_c$ around each spin~\cite{Ogawa15,Khoshlahni19,ViolaKusminskiy2019}, 
\begin{align}
 U(t) &= -\frac{i\theta_F c \sqrt{\epsilon_r} \epsilon_0 a^3}{2\omega} \frac{{\bf M}({\bf r})}{M_s} \cdot [{\bf E}^{*}(t) \times {\bf E}(t)],
\end{align}
the Faraday coupling is $\alpha_F = \theta_F V_c c \sqrt{\epsilon_r}$, where $\theta_F$ is the Faraday angle per unit distance, and ${\bf B}_F(t) = \epsilon_0/(2i\omega) [{\bf E}^{*}(t) \times {\bf E}(t)]$ is the effective optical spin density. The IFE coupling Hamiltonian is then written as
\begin{align}
 H_{\rm IFE} = \alpha_F {\bf B}_F(t) \cdot \sum_i {\bf m}_i.
\end{align}

To estimate the light-matter coupling strength, we write the Faraday angle as $\theta_F = \mathcal{V} B$ where $\mathcal{V}$ is the so-called Verdet constant, which is smaller than $100$ rad/Tm. Taking $E = 10^9$ V/m, $a = 5$ {\AA} and $\lambda = 1240$ nm, giving $\hbar\omega = 1$ eV, we find the characteristic energy scale $g_{\rm IFE} = \alpha_F |{\bf B}_F| = 4.3 \times 10^{-4}$ meV.


\subsection*{Magnetoelectric coupling}
Due to the multiferroic nature of Cu$_2$OSeO$_3$ there is a direct magnetoelectric coupling proceeding with $d-p$-hybridization~\cite{Seki2012}. The resulting magnetoelectric coupling Hamiltonian is
\begin{align}
 H_{\rm m-el} &= - {\bf E} \cdot \sum_i {\bf P}_i \\
 {\bf P}_i &= \gamma (S_i^y S_i^z, S_i^z S_i^x, S_i^x S_i^y). \nonumber
\end{align}
In Refs.~\cite{Seki2012,Liu2013} the coupling constant $\gamma$ is estimated to the value $\gamma = 5.64 \cdot 10^{-27}$ $\mu$Cm by comparison to experiment. This gives, for an electric field of $E = 10^{9}$ Vm$^{-1}$, an interaction energy of $g_{\rm m-el} \approx 10^{-23}$ J or equivalently $g_{\rm m-el} \approx 1$ meV.


\subsection*{Magnon-phonon coupling}
The spin-phonon coupling of Cu$_2$OSeO$_3$ has been discussed in the context of non-reciprocal magnon propagation~\cite{Nomura2019}, where a coupling Hamiltonian of the form
\begin{align}
 H_{\rm s-ph} &= \gamma D \partial_z u_x (S_i^y S_j^z - S_i^z S_j^y) \nonumber \\
&+ \gamma D \partial_z u_y (S_i^z S_j^x - S_i^x S_j^z) \nonumber
\end{align}
was derived for a phonon propagating along the $z$-direction. Assuming acoustic phonons propagating along all cubic axes are excited with the same probability, this gives a spin-phonon coupling Hamiltonian
\begin{align}
 H_{\rm s-ph} &= \gamma {\bf D} \cdot ({\bf S}_i \times {\bf S}_j)
\end{align}
where the DMI vector has been shifted according to the following
\begin{align}
 D_x &= D_{x0}(1 + \gamma [\partial_y + \partial_z] u_x) \\
 D_y &= D_{y0}(1 + \gamma [\partial_z + \partial_x] u_y) \nonumber \\
 D_z &= D_{z0}(1 + \gamma [\partial_x + \partial_y] u_z). \nonumber
\end{align}

To estimate the size of the DMI modulation, we note that in Ref.~\cite{Nomura2019} the value of $\gamma$ was estimated to be in the range $\gamma = 50 - 90$ by fitting the calculated magneto-chiral effect towards experiment. With an estimate of the phonon derivatives of $\partial_j u_i \approx k_j u_i \approx 0.01$, the spin-phonon coupling can still give a modulation of the DMI on the order of 50 \%. This gives a characteristic energy scale of $g_{\rm m-ph} = 0.5D = 5.4$ meV. This large value is in line with previous studies, where the strain-induced modulation of the DMI has been found to be very large for a number of chiral magnets~\cite{Shibata2015,Deng2020}. 

Although the time-dependence of the phonon coordinate $u$ could in principle be obtained by a full dynamical simulation of the coupled vibrational modes of Cu$_2$OSeO$_3$, such a calculation becomes prohibitive in practice due to the complexity of the system. Here we instead take a phenomenological approach and parameterize the time-dependence of the DMI with a log-normal function, whose onset is determined by the pulse parameters of the laser electric field and whose decay is set by the lifetime of the phonon modes.


\end{document}